\documentclass[iop]{emulateapj}
\usepackage{natbib}
\usepackage{booktabs}
\usepackage{appendix}

\newcommand{\citazione}[1]{(\citealt{#1})}

\shortauthors{Zanella et al.}

\begin{document}


\title{The role of quenching time in the evolution of the mass--size
  relation of passive galaxies from the WISP survey\altaffilmark{*}}


\author{A. Zanella\altaffilmark{1,2,3}, C. Scarlata\altaffilmark{1},
  E.M. Corsini\altaffilmark{2,4}, A.G. Bedregal\altaffilmark{5}, E. Dalla
  Bont\`a\altaffilmark{2,4}, H. Atek\altaffilmark{6},
  A. J. Bunker\altaffilmark{7,8}, J . Colbert\altaffilmark{9}, Y. S. Dai\altaffilmark{10}, A. Henry\altaffilmark{11}, M. Malkan\altaffilmark{12}, C. Martin\altaffilmark{13},
  M. Rafelski\altaffilmark{11}, M. J. Rutkowski\altaffilmark{1},
  B. Siana\altaffilmark{14}, 
  H. Teplitz\altaffilmark{15}}

\altaffiltext{*}{Based on observations with the NASA/ESA {\it Hubble Space Telescope}, obtained at the Space Telescope Science Institute, which is operated by AURA, Inc., under NASA contract NAS 5-26555.}
\altaffiltext{1}{Minnesota Institute for Astrophysics, University of Minnesota, Minneapolis MN 55455, USA}
\altaffiltext{2}{Dipartimento di Fisica e Astronomia ``G. Galilei'', Universit\`a di Padova, vicolo dell'Osservatorio 3, 35122 Padova, Italy}
\altaffiltext{3}{Laboratoire AIM, CEA/DSM-CNRS-Universit\'e Paris Diderot, Irfu/Service d'Astrophysique, CEA Saclay, Orme des Merisiers, 91191 Gif-sur-Yvette Cedex, France}
\altaffiltext{4}{INAF - Osservatorio Astronomico di Padova, vicolo dell'Osservatorio 5, 35122, Padova, Italy}
\altaffiltext{5}{Department of Physics and Astronomy, Tufts University, Medford, MA 02155, USA}
\altaffiltext{6}{Spitzer Science Center, Caltech, Pasadena, CA 91125, USA}
\altaffiltext{7}{Department of Physics, University of Oxford, Denys Wilkinson Building, Keble Road, Oxford, OX13RH, United Kingdom}
\altaffiltext{8}{Affiliate Member, Kavli Institute for the Physics and Mathematics of the Universe, 5-1-5 Kashiwanoha, Kashiwa, 277-8583, Japan}
\altaffiltext{9}{Harvard-Smithsonian Center for Astrophysics, 60 Garden Street, Cambridge, MA 02138, USA}
\altaffiltext{10}{Infrared Processing and Analysis Center, 770 South Wilson Avenue, Pasadena, CA 91125}
\altaffiltext{11}{NASA Postdoctoral Program Fellow , Goddard Space Flight Center, Code 665, Greenbelt, MD 20771, USA}
\altaffiltext{12}{Department of Physics \& Astronomy, University of California Los Angeles, Los Angeles, CA 90095, USA}
\altaffiltext{13}{Department of Physics, University of California, Santa Barbara, CA, 93106, USA, cmartin@physics.ucsb.edu}
\altaffiltext{14}{Department of Physics \& Astronomy, University of
  California Riverside, Riverside, CA 92521, USA}
\altaffiltext{15}{Infrared Processing and Analysis Center, Caltech, Pasadena, CA 91125, USA}
%\altaffiltext{9}{Department of Astronomy, Caltech, Pasadena, CA 91125, USA}
%\altaffiltext{10}{Observatories of the Carnegie Institution for Science, Pasadena, CA 91101, USA}
\altaffiltext{**}{Contact email: {\it anita.zanella@cea.fr}}
% \altaffiltext{3}{Carnegie Observatories}
%Supernova Ltd., Olde Yard Village #131, Northsound Road, Virgin Gorda, British Virgin Islands}

%ICRANet, Piazzale della Repubblica 10, I-65100 Pescara, Italy


\begin{abstract}
We analyze how passive galaxies at $z\sim 1.5$ populate the mass--size
  plane as a function of their stellar age, to understand if the
  observed size growth with time can be explained with the
  appearance of larger quenched galaxies at lower redshift. We use a sample of 32 passive
  galaxies extracted from the Wide Field Camera 3 Infrared
Spectroscopic Parallel (WISP) survey with spectroscopic redshift 1.3 $\lesssim z \lesssim$ 2.05, specific star--formation rates lower than $0.01$Gyr$^{-1}$, and stellar masses
  above $4.5\times 10^{10}\, \mathrm{M_{\odot}}$. All galaxies
  have spectrally determined stellar ages from fitting of their
  rest-frame optical spectra and photometry with stellar population
  models. When dividing our sample into young (age $\leq 2.1$ Gyr)
  and old (age $> 2.1$ Gyr) galaxies we do not find a significant trend
  in the distributions of the difference between the observed radius and the one predicted
  by the mass--size relation. This result indicates that the relation between the
  galaxy age and its distance from the mass--size relation, if it
  exists, is rather shallow, with a slope $\alpha \gtrsim -0.6$.
  At face value, this finding suggests that multiple dry and/or wet minor mergers,
  rather than the appearance of newly
  quenched galaxies, are mainly responsible for the observed
  time evolution of the mass--size relation in passive
  galaxies.
 \end{abstract}



\section{Introduction}\label{sec:introduction}
In recent years many efforts have been devoted to observe early-type
galaxies (ETGs) at high redshift to understand how these
objects assembled, evolved, and became quenched. The discovery of a widespread
population of passively evolving ETGs at redshift $z > 1.5$ showed
that the star formation quenching in massive galaxies was already
under way by $z \sim 2$ (e.g.,
\citealt{Mancini2009}). A large fraction of these high redshift
passive galaxies show effective radii between a factor 2 to 5 smaller than local counterparts of comparable stellar masses
\citep[e.g.,][]{Daddi2005}.  This result has been confirmed by several
studies (e.g., \citealt{Trujillo2007}, \citealt{Cimatti2008}, \citealt{Cassata2010}, \citealt{Carollo2013},
\citealt{Vanderwel2014}), and found to be robust
with respect to bias against low surface brightness at high redshifts
(e.g., \citealt{Valentinuzzi2010a}).  In the local Universe, ETGs with similar stellar densities
appear to be quite rare (\citealt{Trujillo2009},
\citealt{Poggianti2012}), although it has been suggested that they
could have survived as the cores of present-day massive spheroids
(\citealt{Hopkins2009}, \citealt{vanDokkum2014}).

This discovery has ignited an important debate. The problem is not the
existence of these compact ETGs: $z\sim 3$ submillimeter
galaxies have comparable masses, sizes and number
density, and have been
identified as their possible precursors
\citep[e.g.,][]{Cimatti2008, Bedregal2012}. The open
issue is how these high-$z$ compact galaxies can evolve to their
present form, inflating their sizes up to a factor of 4, while at the
same time following the tight correlations observed in the local
universe (e.g., the fundamental plane).

Various mechanisms have been suggested to explain the growth of ETGs
with time, although observations are still inconclusive as to which of
them may be favorable. One of the most popular mechanisms invokes the
accretion of multiple small satellites
\citep[e.g.,][]{Naab2009}. These minor mergers leave the mass of the
main galaxy relatively unchanged, while completely disrupting the
satellites through strong tidal interactions. The accretion of
stripped infalling stellar material is expected to increase the
size of the merger remnant, without igniting intense star-formation,
particularly if the satellites do not contain large amounts of gas
(e.g., \citealt{Hopkins2009},
\citealt{Oser2012}). Some observational studies suggest that this
mechanism may account for $\sim$50\% of the apparent size evolution,
at least at redshift $0 < z < 1$ (\citealt{Lopez-Sanjuan2012},
\citealt{Newman2012}). Despite the implications of these observational
results, there is a problem explaining the size evolution with
multiple minor mergers. \citet{Nipoti2012} found that
multiple minor mergers would introduce more scatter than observed in the
low--redshift scaling relations that link the galaxy stellar mass,
effective radius and velocity dispersion, unless the progenitors were already
finely tuned to occupy a very tight region in the mass-radius
plane. Such fine tuning is difficult to explain, and leaves open the
question of \emph{when} and \emph{how} the mass-size relation is first
created. Moreover, \cite{Hopkins2009} highlighted that in the merging
scenario a non-negligible fraction of compact galaxies ($\lesssim$
10\%) should survive to $z \sim 0$, while observations by
\cite{Trujillo2009} show that only 0.03\% of local galaxies have
stellar densities comparable to those of high redshift ETGs.

Adiabatic expansion through significant mass loss can also lead to
size growth \citep{Fan2010}. A galaxy that
loses mass as a result of supernova/AGN--driven winds will adjust its
size in response to the shallower central potential
\citazione{Newman2012}. This mechanism would induce a sort of
``puffing-up'' of the galaxy arising from the loss of baryonic mass,
with an effective size increase. However, the puffing-up only occurs
when the system is highly active and young \citep[in terms of its
stellar population,][]{Ragone2011}, and produces a fast expansion (a
few dynamical times, $\sim 10^8$ yr). Thus one would expect only
a minority of objects to be passive and compact, at odds with observations.

The problem has also been explored from a different perspective (the so called ``progenitor bias'' scenario): instead of 
explaining the evolution of the mass--size relation with the growth of \emph{individual} galaxies with time, it has been suggested that it is the
\emph{population} of ETGs that changes, with larger quenched
galaxies appearing later (\citealt{Saracco2011},
\citealt{Valentinuzzi2010}, \citealt{Cassata2013}). This may be linked to the evolution of the average density in the
Universe due to Hubble expansion, with lower density haloes collapsing later in time than
denser ones (e.g. \citealt{Saracco2011}, \citealt{Carollo2013}). However, the relative importance of the two mechanisms (individual versus population growth)  is still highly controversial (\citealt{Vanderwel2014}, \citealt{Bernardi2010}, \citealt{Cassata2011}, \citealt{Poggianti2013}, \citealt{Belli2015}, \citealt{Keating2015}). If the redshift evolution of the mass--size relation is due to the appearance of newly quenched large galaxies, then one would expect that, at any given mass and time, the larger
galaxies should on average be younger than the smaller ones.  Here we
test this prediction using a sample of $z\sim 1.5$ passive ETGs
observed as part of the Wide Field Camera 3 (WFC3) Infrared
Spectroscopic Parallel (WISP) survey \citazione{Atek2010}.

Throughout the paper we assume a flat cosmology with $H_0 = 70$ km
$\mathrm{s^{-1} \, Mpc^{-1}}$, $\Omega_M = 0.3$, and
$\Omega_{\Lambda} = 0.7$. Photometric magnitudes are expressed in the
AB system \citazione{Oke1983}.

\section{Observations and data analysis}\label{sec:data}
The sample presented in this work includes 34 passive galaxies
identified in the WISP survey, a pure-parallel Hubble Space Telescope
(\textit{HST}) program to obtain near-infrared slitless spectra together with optical and infrared (IR)
imaging of hundreds of independent fields in the sky. The data have
been presented in detail in \citet{Atek2010}. Briefly, we consider
here the first 27 fields observed with both \textit{HST} WFC3 grisms (G102, and
G141; with resolving power $R=210$ and 130, respectively) as well as
with the WFC3--UVIS camera in the optical.  The IR spectra cover the
wavelength range between $0.85\le \lambda \le 1.6\mu$m, with
approximately 0.1$\mu$m overlap between the two grisms that allows us,
together with the IR imaging, to
check for proper photometric calibration and sky subtraction. The data
were reduced and the spectra extracted with a combination of a custom
pipeline described in \citet{Atek2010} and the aXe software
\citep{Kummel2009}. In addition, we implemented a new cleaning
algorithm to properly account for contamination from
overlapping spectra \citep[see details
in][]{Bedregal2012}. The imaging was obtained
with the F475X and F600LP (apart from the two deepest fields that were
observed with F606W and F814W instead), and with the F110W and F160W
filters, respectively \citep[see][]{Bedregal2012}.
 
\citet{Bedregal2012} studied the properties of a sample of
$H<23$ mag galaxies preselected on the basis of their $J-H$
color. They measured spectroscopic redshifts and stellar population
properties of the galaxies by simultaneously fitting the broad band
photometric points and spectra. Here we present the size measurements
for the subsample of passive galaxies selected to have a specific
star formation rate sSFR $\lesssim$ 0.01 Gyr$^{-1}$, redshift $z>
1.3$ and $M_* > 4.5 \times 10^{10}\, \mathrm{M_{\odot}}$. This is the minimum stellar mass
measurable at $z \sim 1.5$ (the average redshift of the sample), for a maximally old stellar population
model. These limits have been chosen to select a sample of massive and
passive galaxies and their robustness is discussed in
\citet{Bedregal2012}. If we were to consider a more conservative
sample selection based on the minimum stellar mass measurable at $z \sim 2$
(the highest redshift in our sample) $M_* > 7.9 \times 10^{10}\,
\mathrm{M_{\odot}}$, our results would not change substantially (see
Section 4). We  verified that the stellar mass
function of our galaxy sample is consistent with the one
determined by \citet{Muzzin2013}, for quiescent galaxies, at the same
redshift. All galaxies have accurate luminosity-weighted stellar ages derived fitting the grism spectra with stellar population models. Using the same set of simulations performed by Bedregal et al. (2013) we find that for our sample, the stellar ages are recovered with an accuracy of 35\%. 

\section{SIZE ANALYSIS}
\label{sec:morphol}
For the structural analysis of the light distribution of our
sample galaxies we use the deeper $J_{\mathrm{110}}$ images. The sky background has been previously subtracted as discussed in \citet{Atek2010} and Colbert et al. (2013).  
We perform the measurements with two different fitting algorithms: the widely used
GALFIT code \citep{Peng2010} and the alternative GASP2D code \citep{Mendezabreu2008}.  To be consistent with previous works in the
field, we fit the galaxy light distribution with a single S\'ersic
law \citazione{Sersic1968}.

Both codes require as input the instrumental Point--Spread--Function
(PSF). Our data were taken in parallel to observations performed
with the Cosmic Origin Spectrograph and Space Telescope Imaging Spectrograph, and no spatial dithering
was done in between different exposures. Because of this,
the final PSF is undersampled at the pixel size (0\farcs13 pixel$^{-1}$) of the WFC3--IR
camera.  We provide both GALFIT and GASP2D with a PSF that we obtained as
the median of 18 unsaturated stars across the 27 analyzed fields.  We used this median 
PSF to fit individual stars in each field, and found residuals of at most 20\%, irrespective of the field. Before fitting the
galaxies, we masked any foreground and background sources, as well as
detector artifacts that can contaminate the surface brightness
distribution. The main differences between the two codes is in the way
the initial values for the parameters are determined. GALFIT does not
provide a way to estimate them, while GASP2D internally determines
their values by performing a fit on the one--dimensional surface
brightness profile obtained using the IRAF task \texttt{ellipse} (see
\citealt{Mendezabreu2008} for details). We visually inspected
  all the residuals to check for the reliability of the fits and
  we find that for each sample galaxy they are lower than 20\%.

The galaxy effective radii ($r_{\mathrm{e}}$) measured with the two algorithms are
consistent within the uncertainties. In the following analysis we use
the sizes determined by GASP2D, but our conclusions would not
change if we switch to GALFIT instead.
For each galaxy we compute the circularized effective radius as
$r_{\mathrm{e}}^{\mathrm{circ}} = r_{\mathrm{e}}\sqrt{q}$, where $q$
is the galaxy axial ratio. 

\vspace{0.5cm} We determine the uncertainties associated to the $r_{\mathrm{e}}$
measurements through Monte Carlo simulations. We created 1000
artificial galaxies with S\'ersic parameters randomly chosen in the range of values
observed for real galaxies (total magnitude $19 \leq m_{\mathrm{tot}}
\leq 24$, effective radius $0.1" \leq r_{\mathrm{e}} \leq 1.5"$,
S\'ersic index $0.5 \leq n \leq 12$, axial ratio $0.2 \leq q \leq 1$,
position angle $0^\circ \leq PA \leq 180^\circ$).  All the models were
convolved with the PSF image and we added Poisson noise to reproduce
the observations. The best--fit S\'ersic parameters were then derived
using both GALFIT and GASP2D.  

For each parameter, we estimated the fractional uncertainty as:
\begin{equation}
\epsilon = \frac{p_{\mathrm{out}} - p_{\mathrm{in}}}{p_{\mathrm{out}}}
\label{eq:relative_errors}
\end{equation}

\noindent
where $p_{\mathrm{out}}$ is the fitted parameter and $p_{\mathrm{in}}$
is the input value.  We then computed the median and the 16th and 84th
percentiles in bins of $p_{\mathrm{out}}$, to estimate
the systematics together with upper and lower uncertainties on the parameters. We excluded from the sample two galaxies with output effective radii smaller than $1$ kpc since we believe that $r_{\mathrm{e}} < 1$ kpc values are not reliable,
and we therefore set $1$ kpc as the minimum size we are able to
resolve. With Monte Carlo simulations we also checked that galaxies
with $r_{\mathrm{e}} < 1$ kpc do not enter the sample with an
overestimated size. This limit is higher than similar depth surveys
performed with WFC3: it results from the lack of spatial dithering
between exposures. With simulations we
quantify the impact of this size limit on our results (Section 4).
The uncertainty  associated with the circularized radius is calculated as $\epsilon_{r_{\mathrm{e}}}^{\mathrm{circ}} =
\epsilon_{r_{\mathrm{e}}}\sqrt{q}$, because we find that the axial ratio uncertainty is negligible.

\vspace{0.5cm} The S\'ersic  best--fit circularized half--light
radius, S\'ersic index, stellar mass, stellar age and redshift for each galaxy in the final sample
are presented in Table \ref{tab:final_sample}. We notice that half of
the galaxies in our sample have a S\'ersic index $n < 2.5$, typically
associated to disk-dominated galaxies, in agreement with, e.g.,
\cite{vanderWel2011} and N14.

\begin{table}[t!]
\centering
\footnotesize
\caption{Circularized effective radius, S\'ersic index, stellar mass, age, and redshift of the galaxies in the final sample.}
\label{tab:final_sample}
\begin{tabular}{p{1.8cm} c c c c c}
\\
\toprule
\midrule
Galaxy &$r_{\mathrm{e}}^{\mathrm{circ}}$ & $n$ & $\log(M_{\star}/\mathrm{M_{\odot}})$  & Age & $z$\\
       &     (kpc)             &            &      &     (Gyr)        &  \\
(1) & (2) & (3) & (4) & (5) & (6)\\
\midrule
Par66\,ID135  &   1.33 $\pm$ 0.35 & 2.24 & 11.24  & 2.00  & 1.80 \\
Par67\,ID108 &    1.05 $\pm$ 0.33 & 1.07 & 10.71     &  1.02   & 1.35  \\
Par67\,ID140   &  1.75 $\pm$ 0.38 & 0.72 & 11.30 &  2.75 &  2.05 \\	 
Par67\,ID82    &   1.36  $\pm$ 0.26 & 2.29 & 11.08  &  4.00  & 1.35 \\	 
Par73\,ID152   &  2.71  $\pm$ 1.19 & 1.23 & 10.74  &  2.75 & 1.50 \\
Par73\,ID47    &  2.74  $\pm$ 0.17 & 4.03 & 11.16  &  0.90 &1.45 \\ 	 
Par73\,ID57    &  1.67  $\pm$ 0.51 & 1.35 & 11.22 &  0.90  & 1.60 \\ 	 
Par74\,ID37    &  2.54 $\pm$ 0.82 & 1.72 & 11.44  & 2.50 & 1.60 \\	 
Par76\,ID26    &  2.32 $\pm$ 1.02 & 7.38 & 11.77  & 4.25 &1.40 \\
Par76\,ID41    &  2.29  $\pm$ 0.84 & 0.74 & 11.26 & 4.75 & 1.35 \\	 
Par76\,ID60    &  2.96  $\pm$ 1.15 & 3.54 & 11.73   & 3.50 &1.70 \\	 
Par76\,ID62   &   1.64 $\pm$ 0.59   & 17.53 & 11.50  &  3.50 & 1.70  \\
Par76\,ID77   &  1.23  $\pm$ 0.37 & 0.69 & 11.31 & 0.45  & 2.05 \\	 
Par79\,ID19    &  3.06  $\pm$ 0.70 & 7.92 & 11.55  &  2.20 & 1.35 \\	 
Par79\,ID86    &  1.53  $\pm$ 0.61 & 8.76 & 11.05  &  1.02 & 1.90 \\
Par80\,ID28    &  5.04  $\pm$ 0.56 & 4.15 & 11.43  &  4.00 & 1.40 \\	 
Par80\,ID35   &  1.73 $\pm$ 0.54 & 3.33 & 11.14 & 1.61 & 1.55 \\	 
Par80\,ID50    &  2.19 $\pm$ 0.51 & 1.10 & 11.17  &  3.00 & 1.40 \\	 
Par80\,ID93    &  4.13 $\pm$ 0.79 & 2.92 & 11.04 &  3.50  & 1.85 \\	 
Par84\,ID57    &  1.83  $\pm$ 0.57 & 2.09 & 11.43  &  4.00 &1.45 \\	 
Par87\,ID118  &   3.14  $\pm$ 0.40& 5.14 & 11.38   &  2.20 & 1.70 \\	
Par87\,ID125     &   1.34 $\pm$ 0.53  & 7.29 &  10.66 & 0.40  &   1.85  \\ 
Par87\,ID54    &  1.28 $\pm$ 0.35   & 1.25 & 11.34  &  3.00  & 1.50  \\
Par87\,ID87    &   1.50 $\pm$ 0.53 & 6.05 & 11.12 & 2.00  & 1.65 \\	 
Par87\,ID95    &  2.73 $\pm$ 0.35 & 0.83 & 10.71 & 1.02 & 1.60 \\	 
Par96\,ID62    &  1.28 $\pm$ 0.35 & 2.47 & 10.91 & 1.02  & 1.75 \\	 
Par115\,ID83   &   2.04 $\pm$ 0.45 & 2.66 & 11.01 & 1.80  & 1.65 \\	 
Par120\,ID64   &   1.74  $\pm$ 0.54 & 3.43 & 11.23 & 1.61  & 1.50\\	 
Par120\,ID84   &   4.01  $\pm$ 1.60 & 2.21 & 11.28 &  2.20  & 1.65 \\	 
Par136\,ID55   &   2.79  $\pm$ 1.22 & 2.11 & 10.68  & 0.90 & 1.65 \\
Par136\,ID77   &   3.55  $\pm$ 0.62 & 0.90 & 10.84 & 1.14  & 1.65 \\	 
Par147\,ID46   &   2.38  $\pm$ 1.04 & 2.52 & 11.02  & 0.64 & 1.46\\
\bottomrule
\\
\begin{minipage}{8cm}
\scriptsize
NOTES -  Col. (1): galaxy name defined as in \citet{Bedregal2012}. Col. (2): circularized effective radius with associated relative error. Col. (3): S\'ersic index. Col. (4): logarithm of the stellar mass. Col. (5): age. Col. (6): redshift. \\
\end{minipage}
\end{tabular}
\end{table}
\normalsize

\section{Results and Discussion}
\label{sec:discussion}
\begin{figure}[t!]
\centering
\includegraphics[width=0.5\textwidth]{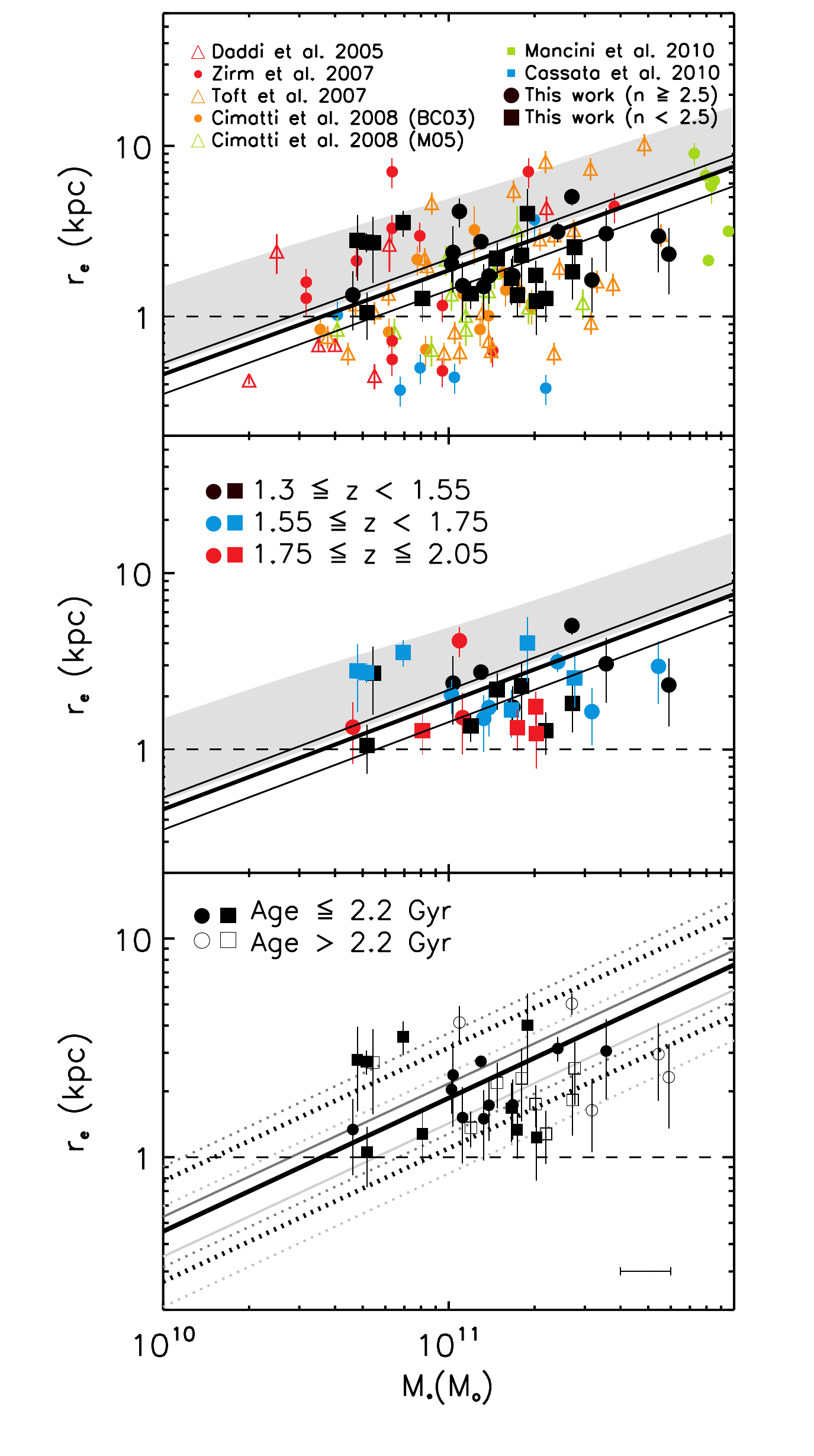}
\caption{The mass-size relation measured with the WISP data at $z \sim
  1.5$. We show circularized effective radius vs stellar mass for the
  present sample (black symbols), with a distinction based on the S\'ersic index (circles and squares), and compare the results drawn from the
  literature. \emph{Top}: Circularized effective radius versus stellar
  mass. The local mass-size relation and its scatter \citep[from
  ][]{Shen2003} are shown, for illustrative purposes
  as a grey band. The solid thick line shows
  the best--fit mass--size relation from N14, computed
  at $z=1.5$, while the thin lines were computed at the lowest (1.35) and
  highest (2.05) redshifts of our sample galaxies. All the galaxy radii shown here are circularized, except for those by \citet{Mancini2009}. \emph{Middle}: Same as top panel, but showing our sample in three groups of redshift (black,  blue, and red symbols). Again we separate galaxies with low and high S\'ersic index (circles and squares). \emph{Bottom}:
 Same as the top panel, but showing our sample in two groups of
 stellar age (filled and empty symbols). In the bottom right corner we
  show the median uncertainty in stellar mass for our sample of
  galaxies \citep{Bedregal2012}. Our effective radii measurement
  limits are shown (dashed line).}
\label{fig:comparison_literature}
\end{figure}

We compare the stellar mass and size of our galaxies with those in
the literature in the top panel of
Figure~\ref{fig:comparison_literature}. Before placing the literature
data on the mass--size plane, we homogenized all the masses to
Salpeter initial mass function (IMF,
\citealt{Salpeter1955})\footnote{The scaling factors between the
  \citet{Chabrier2003}, \citet{Kroupa2001} and \citet{Salpeter1955} IMFs that we adopted are: $\log(M_{\mathrm{Chabrier}}) = \log(M_{\mathrm{Kroupa}}) - 0.04$ \citazione{Cimatti2008}  and $\log(M_{\mathrm{Chabrier}}) = \log(M_{\mathrm{Salpeter}}) -0.25$ \citazione{Salimbeni2009}} and corrected them for the systematics implied by the different ages of the adopted synthetic
stellar population models (SPMs). We scaled all masses to the
\cite{Bruzual2003} SPMs adopting the \cite{Salimbeni2009} relations
$\log(M_{\mathrm{M05}}) \simeq \log(M_{\mathrm{CB07}})$ and
$\log(M_{\mathrm{M05, CB07}}) = \log(M_{\mathrm{BC03}}) - 0.2$ where
BC03, M05, and CB07 indicate \cite{Bruzual2003}, \cite{Maraston2005},
and \cite{Charlot2007} SPMs, respectively.

We limit the comparison to only those works where the sample selection
is based on the galaxy specific star--formation rates and that span
a similar redshift range to ours. The only exception is the \cite{Mancini2009}
sample, where a morphological selection criterion (based on the
S\'ersic index) was also
applied. Figure~\ref{fig:comparison_literature} shows that our
measurements are consistent with the results found at similar
redshifts by other works in different fields, once the size
lower-limit (dashed horizontal line) is considered. The thick solid
line shows the best--fit mass--size relation derived by
\citet[N14 hereafter]{Newman2014} for field galaxies, computed at the median
redshift of the sample. The thin lines were computed at the lowest and
highest redshifts of our sample galaxies. 

%The question we want to investigate in this paper is whether the
%evolution of the mass--size relation for passive  galaxies is due to the later 
%quenching of large galaxies, compared to compact ones 
%\citep[e.g.,][]{vanderwel2008,saracco2014, tacchella2015}.  
%
%In fact, \citet{Carollo2013} and \citet{Cassata2013} recently proposed that, at a given time and for a given mass, the larger galaxies
%among the passive objects should have been quenched relatively more
%recently than the more compact ones, but our results seem not to support this scenario. 

\begin{figure}[t!]
\centering
\includegraphics[width=0.5\textwidth]{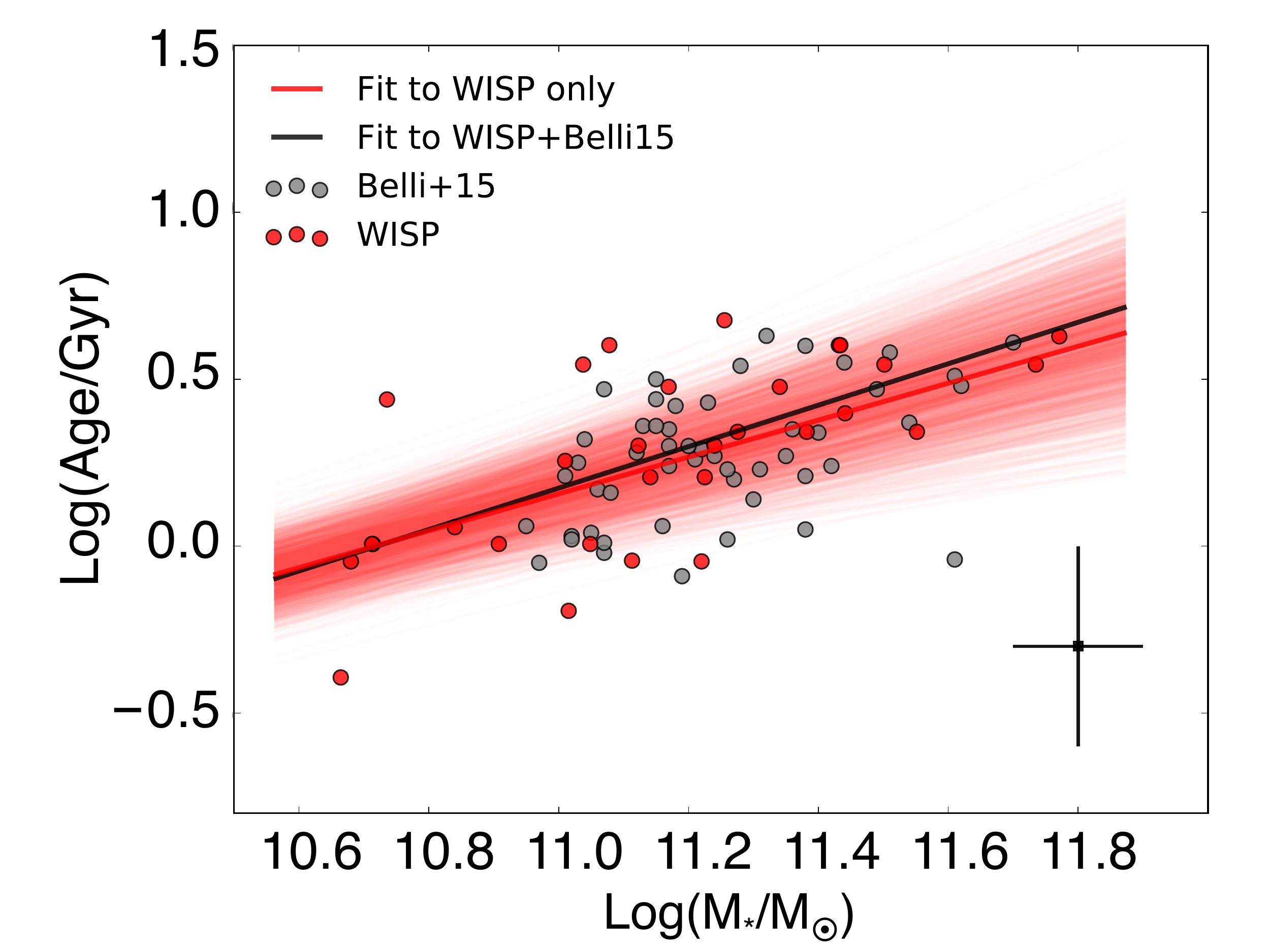}
\caption{Stellar age and mass of quenched galaxies are
    strongly correlated.  Stellar age as a function of stellar mass
    for the WISP galaxies (red points) and Belli et al. (2015) sample
    (gray points). Belli et al. (2015) stellar masses were converted
    to Salpeter IMF following Salimbeni et al. (2009).  The best--fit
    lines to the WISP sample alone and to the combined WISP and Belli
    et al. (2015) samples are shown with red and black lines, respectively. The red band shows the uncertainty on the fit to the WISP data alone (the uncertainty on the fit to the combined sample is similar, and not shown for clarity).}
\label{fig:age_mass}
\end{figure}

In the bottom panel of Figure~\ref{fig:comparison_literature} we reproduce the mass--size
relation for our galaxies, dividing the sample in old (stellar
age $> 2.1$ Gyr) and young objects (age $\leq 2.1$ Gyr). The age
separation was chosen to divide the galaxies in two similar size
samples. Figure~\ref{fig:comparison_literature}  shows how the most massive galaxies
($M_* > 2 \times 10^{11}\, \mathrm{M_{\odot}}$) tend to be older than less massive ones, a trend compatible 
 with other observational results indicating that more massive
 galaxies form the bulk of their stars earlier
 \citep[e.g.,][]{Thomas2005, Kaviraj2013}. We quantify this trend in Figure~\ref{fig:age_mass}, where we show the stellar age versus the 
 stellar mass, for the galaxies in our sample as well as for a sample of similarly selected objects identified by Belli et al. (2015). 
 The correlation between the stellar population age and mass has a Spearman correlation coefficient of 0.66, which has  a 
 probability of $10^{-5}$ of  resulting by chance. At any stellar
 mass, the stellar age ($A_{\mathrm{M}}$) can be expressed as: $\log{(A_{\mathrm{M}}/\rm{Gyr})}=(0.55\pm0.09)\log{M_{10.5}} 
 - (0.12\pm 0.07) $,  where $M_{10.5}$ is the stellar mass in units of $10^{10.5}$M$_{\odot}$.
 
\begin{figure}[t!]
\centering
\includegraphics[width=0.5\textwidth]{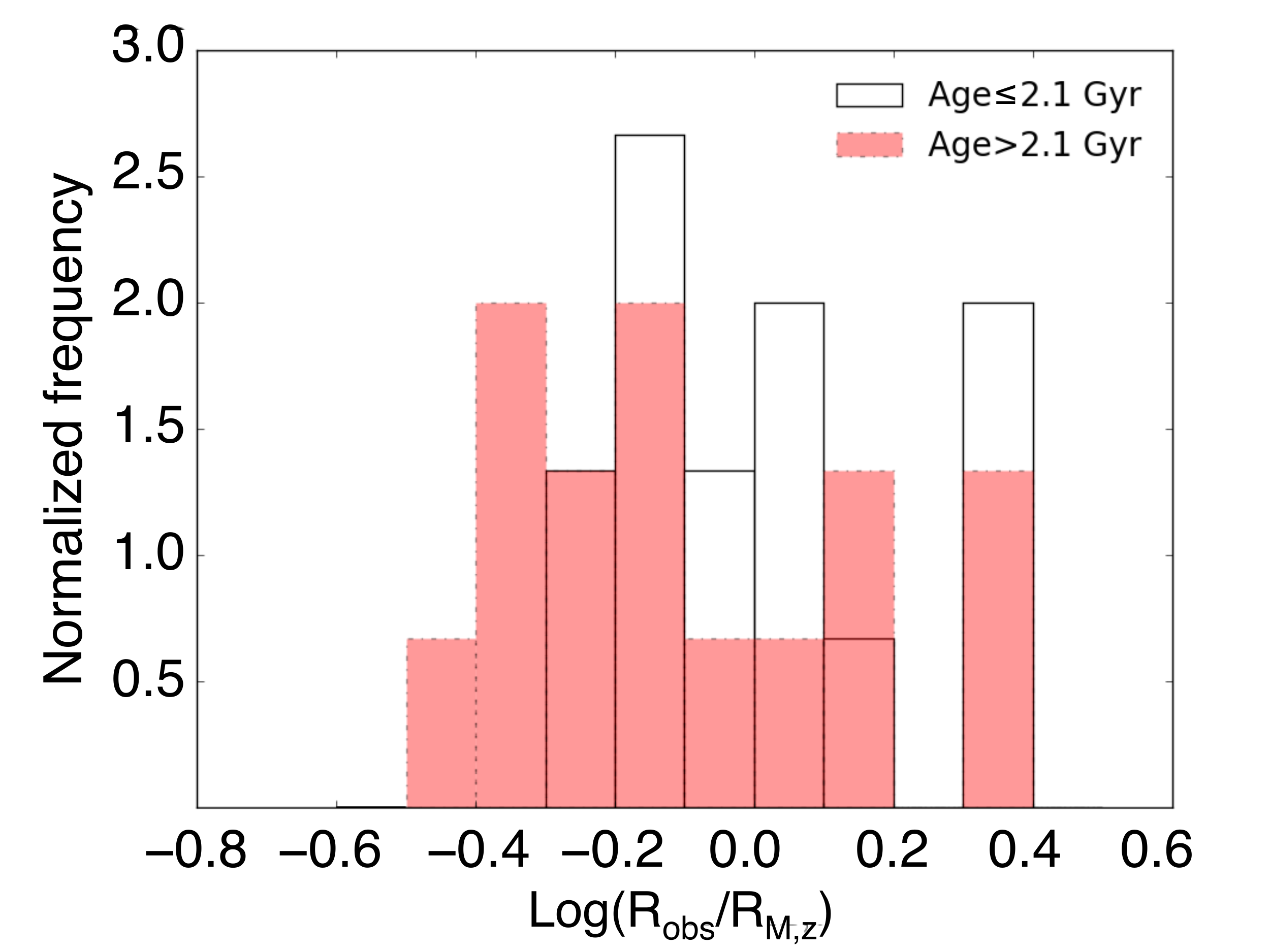}
\caption{Distribution of $\Delta_{\mathrm{lr}}$   for the WISP passive galaxies. Although the distributions of young
  and old galaxies have different median values, they are not
  statistically different. Pink and white lines represent old and young
  galaxies, respectively (see text for details).}
\label{fig:deviation_r}
\end{figure}

To quantify whether a trend between the stellar age and the deviation from the $z\sim 1.5$ mass-size
relation exists we compute for each galaxy the parameter
$\Delta_{\mathrm{lr}}= \rm{log}(\it{R_\mathrm{obs}/R_{\mathrm{M,z}}})$, i.e., the vertical difference between the observed galaxy 
size ($R_{\mathrm{obs}}$) and the size expected from the galaxy's redshift and stellar mass 
($R_{\mathrm{M,z}}$), using the N14 mass--size relation. Values of
$\Delta_\mathrm{lr}>0$ ($<0$) indicate 
that galaxies are above (below) the mass--size relation at the galaxy redshift. The distributions of  $\Delta_\mathrm{lr}$ 
for the old and young galaxies with $M_* > 4.5 \times
10^{10}\, \mathrm{M_{\odot}}$ are shown in Figure~\ref{fig:deviation_r}. The  medians of the two distributions are 
$-0.02^{+0.36}_{-0.16} $ and $-0.13^{+0.31}_{-0.21} $, for the young
and old samples, respectively (the upper and lower range  show the
84$^{th}$ and 16$^{th}$ percentiles).  The result of a two-sample Kolmogorov--Smirnoff (KS) test  ($D_\mathrm{KS}=0.3$ and $p=0.3$), however, indicates that we cannot exclude that the two samples are 
drawn from the same parent distribution and thus the observed age
difference is not significant. More conservatively considering only galaxies
with $M_* > 7.9 \times 10^{10}\, \mathrm{M_{\odot}}$ we obtain
consistent results (the medians of the young and old galaxy
distributions of $\Delta_\mathrm{lr}$ are $-0.10^{+0.09}_{-0.16}$ and $-0.18^{+0.17}_{-0.14}$, respectively).

The size measurement limit ($r_{\mathrm{e}}>1$ kpc), together with the observed 
dependency between the galaxy stellar age and mass may introduce a bias, particularly 
at the smallest masses, where  our galaxies tend to populate the mass--size plane above 
the best--fit relation derived  at similar redshifts (N14).
To test to what extent we can detect with our data a possible trend of the distance
from the best--fit mass--size relation with age, we performed a
simulation that accounts for both the size selection bias and the
observed mass--age trend. We generated 1000 samples of 32 galaxies with masses and half--light 
radii distributed according to the field mass--size relation and its scatter, determined by N14 at $z\sim 1.5$.
To each galaxy we assign an age ($A$) that depends on its stellar mass
and the distance to the mass--size relation, such that $A =
A_\mathrm{M} + \alpha \Delta_\mathrm{lr}$. We randomize 
the simulated ages and stellar masses according to the typical uncertainties of our observations. 
We then apply the WISP survey limits to the simulated galaxy samples
(i.e., $M_* > 4.5\times 10^{10}\, \mathrm{M_{\odot}}$, and $r_{\mathrm{e}} \ge 1$ kpc), and
recompute the distributions of $\Delta_\mathrm{lr}$ for the subsamples of old and young galaxies. 
For each of the 1000 samples we performed the same analysis described above, 
and compute the KS test between the distributions of $\Delta_\mathrm{lr}$ for the young and old subsamples.  
For $\alpha = 0$ (i.e., no correlation), we find that only 10\% of the
simulated samples show a detectable differences between the old and
young galaxies due to our size limit and the correlation between the
age and stellar mass. We consider decreasing values of $\alpha$ in steps of 0.1, from zero to the $\alpha$ that produces distinguishable distributions. 
Our simulations show that for any $\alpha < -0.64$ we would be able to recover the difference between old and young population at 95\% confidence level in more than 85\% of the simulated
samples. Repeating the same simulations applying a mass limit
  $M_* > 7.9\times 10^{10}\, \mathrm{M_{\odot}}$ we find $\alpha < -0.7$. Our
data, therefore, suggest that the relation between the galaxy age and
its distance from the mass--size relation, if it exists, must have
$\alpha > -0.64$. Performing these simulations considering the
  galaxies formation redshift ($z_\mathrm{f}$)
instead of age, given the observed redshift and the current
cosmology, we conclude that, if a relation $R \sim
(1 + z_\mathrm{f})^\beta$ exists, then it must be $\beta > -0.67$,
otherwise we would have detected the correlation with our current sample.

\section{Conclusions}
\label{sec:summary}

We studied the mass--size relation of a sample of 32
passive galaxies at $z \sim 1.5$ selected from the WISP survey
to have sSFR $< 0.01\, Gyr^1$ and $M_* > 4.5\times 10^{10}\, \mathrm{M_{\odot}}$. All galaxies have accurately
determined stellar ages from fitting the galaxy rest-frame optical
spectra with stellar population models \citep{Bedregal2012}. We
investigate whether younger galaxies have preferentially larger sizes
than older ones with the same stellar mass. Such an observation would indicate that the
mass-size relation evolves due to the appearance of newly quenched large objects in
passive samples.
 
Dividing our sample in young and old galaxies we find no
significant difference in the distributions of $\Delta_{\mathrm{lr}}$, suggesting that the appearance of
newly quenched galaxies may not be the dominant mechanism for the
evolution of the mass--size relation. Our
simulations also indicate that, if a relation exists between the
galaxy age and the distance to the mass--size relation, it has a slope
$\alpha > -0.64$, otherwise we would have detected it. It translates
into a slope of the galaxies size-formation redshift relation $\beta >
-0.67$, given the current cosmology. If we consider in our analysis more conservatively only galaxies with $M_* > 7.9\times 10^{10}\,
  \mathrm{M_{\odot}}$ we obtain consistent results.

Our results suggest that the evolution of the mass--size relation of 
quiescent galaxies is mainly due to the physical growth of individual sources.
Recently \cite{Belli2015} have found  that progenitor bias can explain 
half of the size growth of compact ETGs and that the remaining observed size evolution 
arises from a genuine growth of individual galaxies. The discrepancy is likely due to the fact that they include in the sample ``green valley'' sources, with sSFR  $< 0.1\, \mathrm{Gyr^{-1}}$, while we limit the analysis to those galaxies with  
sSFR $< 0.01\, \mathrm{Gyr^{-1}}$. In fact, it is exactly these sources with higher sSFR that drive  the correlation between 
the age and the size evolution \citep[see Figure~9
in][]{Belli2015}. Our finding is in contradiction with works implying a slope $\beta
  \sim -1$ for the size-formation redshift relation and suggesting that galaxies size scale as the density of the Universe at the time when they formed  (e.g. \citealt{Saracco2011},
  \citealt{Cassata2013}, \citealt{Carollo2013}).
Our results are instead in agreement with the ones by \cite{Trujillo2011} and \cite{Whitaker2012} 
which do not see any age segregation depending on the galaxy size. \cite{Sonnenfeld2014} suggest that the observed size growth can not be explained  
with models invoking  only dry merger, because they would result in a strong flattening of the mass 
density profile with time. This flattening is not observed in the samples of strong lenses for which the total 
mass--density profile could be constrained  \citep{Sonnenfeld2014}. 
The size growth could instead be due to a combination of dry and wet minor mergers: the outer regions of 
massive ETGs could grow by accretion of stars and dark matter, while a small amount of nuclear star formation could keep 
the mass density profile constant with time (e.g., Rutkowski et al. 2014).

%We thus conclude that our results 
%seem to be consistent with this scenario, indicating that at $1.3 \lesssim z \lesssim 2$ the size evolution of passive 
%galaxies is mostly driven by dry and wet minor mergers.

\acknowledgments
We thank the referee for her/his constructive comments which improved the analysis of the results.
We thank Francesco Valentino, Emeric Le Floc'h and Emanuele Daddi for
useful discussions.
EMC and EDB are supported by Padua University through grants 60A02-5857/13, 60A02-5833/14, 60A02-4434/15, and CPDA133894.

%\begin{thebibliography}{}
%\bibitem[Auri\`ere(1982)]{aur82} Auri\`ere, M.  1982, \aap,  109, 301
%\bibitem[Treu et al.(2003)]{treu03} Treu, T. et al., 2003, \apj, 591, 53
%\end{thebibliography}
\bibliographystyle{apj} 
\bibliography{biblio}
\clearpage

\end{document}